\documentclass[letterpaper, 10 pt, conference]{ieeeconf}  %

\IEEEoverridecommandlockouts                              %
\let\labelindent\relax %

\usepackage[colorinlistoftodos,bordercolor=orange,backgroundcolor=orange!20,linecolor=orange,textsize=scriptsize]{todonotes}

\usepackage[english]{babel}
\usepackage[T1]{fontenc}
\usepackage{xargs}
\usepackage{amsmath}
\usepackage{amsfonts}
\usepackage{amssymb}
\let\proof\relax

 \usepackage{amsthm}
  
\usepackage{float}
\usepackage{adjustbox}
\usepackage{bm}
\usepackage{bbm}
\usepackage{dsfont}
\makeatletter
\let\NAT@parse\undefined
\makeatother
\usepackage[numbers,sort&compress]{natbib}

\usepackage{enumitem}

\usepackage[framemethod=tikz]{mdframed}

\newif\ifproofTerm
\proofTermfalse

\usepackage{color}
\definecolor{myblue}{rgb}{0.122, 0.435, 0.698}
\newmdenv[innerlinewidth=0.5pt, roundcorner=4pt,linecolor=myblue,innerleftmargin=6pt,
innerrightmargin=6pt,innertopmargin=6pt,innerbottommargin=6pt]{bluebox}
\definecolor{myred}{rgb}{0.8, 0.1, 0.1}
\newmdenv[innerlinewidth=0.5pt, roundcorner=4pt,linecolor=myred,innerleftmargin=6pt,
innerrightmargin=6pt,innertopmargin=6pt,innerbottommargin=6pt]{redbox}

\usepackage{cleveref}

\Crefname{ALC@unique}{Line}{Lines}

\usepackage{phdalgo}

\newtheorem{theorem}{Theorem}

\newtheorem{proposition}{Proposition}
\newtheorem{lemma}{Lemma}
\newtheorem{remark}{Remark}

\Crefname{corollary}{Cor.}{Cors.}
\Crefname{equation}{Eq.}{Eqs.}
\Crefname{figure}{Fig.}{Figs.}
\Crefname{tabular}{Tab.}{Tabs.}
\Crefname{table}{Tab.}{Tabs.}
\Crefname{theorem}{Thm.}{Thms.}
\Crefname{definition}{Def.}{Defs.}
\Crefname{section}{Sec.}{Secs.}
\Crefname{proposition}{Prop.}{Props.}
\Crefname{assumption}{Asm.}{Asms.}
\Crefname{example}{Ex.}{Exs.}
\Crefname{algocf}{Algo.}{Algorithms}

\Crefname{appsec}{Appendix}{Appendices}

\Crefname{claim}{Claim}{Claims}

\usepackage{algorithm}
\usepackage{algpseudocode}

\newcommand{\norm}[1]{\left\Vert #1\right\Vert}
\newcommand{\normop}[1]{\left\Vert| #1|\right\Vert}
\newcommand{\dssum}{\displaystyle\sum}
\newcommand{\hh}{\hspace{-2pt}}

\graphicspath{{../images/},{../../../SIMU/plots/keep/CDC_Disag},{../../SIMU/plots/keep/smartGridComm/}}
\DeclareGraphicsExtensions{.pdf,.jpeg,.png,.svg}

\newcommand{\eqDef}{\overset{\mathrm{def}}{=}}

\newcommand{\x}{\ell} %
\newcommand{\xx}{\bm{\ell}}
\newcommand{\N}{\mathcal{N}} %
\newcommand{\T}{\mathcal{T}} %
\renewcommand{\H}{\mathcal{H}} %
\newcommand{\X}{\mathcal{X}} %

\newcommand{\lnu}{\underline{\nu}} %

\newcommand{\rr}{\mathbb{R}} %

\newcommand{\Y}{\mathcal{Y}}
\newcommand{\y}{y}
\newcommand{\yy}{\bm{y}}

\renewcommand{\i}{n}
\newcommand{\h}{t}
\newcommand{\ih}{_{\i,\h}}
\renewcommand{\x}{x}
\renewcommand{\xx}{\bm{\x}}
\newcommand{\B}{\mathcal{B}}
\renewcommand{\ss}{\bm{s}}
\newcommand{\lbx}{\underline{x}}
\newcommand{\ubx}{\overline{x}}

\newcommand{\pp}{\bm{p}}
\renewcommand{\P}{\mathcal{P}}

\newcommand{\mc}{\mathcal}

\newcommand{\mr}{\mathrm}

\newcommand{\ul}{\underline}
\newcommand{\ol}{\overline}

\newcommand{\eqd}{\eqDef}
\newcommand{\dsone}{\mathds{1}}
\newcommand{\dsoneN}{\mathds{1}_{\hspace{-2pt}N }  }
\newcommand{\dsoneT}{\mathds{1}_{\hspace{-2pt}T }  }
\newcommand{\tr}{ {\hspace{-2pt}\top } }
\newcommand{\EE}{\bm{E}}

\newcommand{\nt}{_{n,t}}

\newcommand{\Ti}{T}
\newcommand{\Ni}{N}

\newcommand{\llbx}{\bm{\lbx}}
\newcommand{\uubx}{\bm{\ubx}}

\newcommand{\kexp}{^{(k)}}
\newcommand{\xE}{\xx} %
\newcommand{\xEinf}{\xx^\infty}%

\newcommand{\xP}{\yy}%
\newcommand{\xPinf}{\yy^\infty}%
\newcommand{\txt}{\textstyle}
\newcommand{\U}{\mathcal{U}}

\newcommand{\oomu}{\bm{\overline{\mu}}}
\newcommand{\uumu}{\bm{\underline{\mu}}}
\renewcommand{\H}{\T}

\newcommand{\A}{\mr{A}} %

\newcommand{\nnu}{\bm{\nu}}
\newcommand{\mrin}{_{\mr{in}}}

 \renewcommand{\SS}{\bm{S}}

\newcommand{\rhoN}{\rho_{NT}}

\title{\LARGE \bf
A Privacy-preserving Disaggregation Algorithm %
\\ 
for Non-intrusive Management of Flexible Energy %
}

\author{Paulin Jacquot, Olivier Beaude, Pascal Benchimol, St\'ephane Gaubert, Nadia Oudjane%
\thanks{P. Jacquot, O. Beaude, P. Benchimol and N. Oudjane are with EDF R\&D, OSIRIS, Palaiseau, France.}
\thanks{P. Jacquot and S. Gaubert are with Inria Saclay and CMAP, Ecole polytechnique, Palaiseau, France.}
\thanks{{\tt\small paulin.jacquot@polytechnique.edu}}%
}

\makeatletter
\providecommand\theHALG@line{\thealgorithm.\arabic{ALG@line}}
\makeatother

\begin{document}

\maketitle
\thispagestyle{empty}
\pagestyle{empty}

\begin{abstract}
  We consider a resource allocation problem involving a large number of agents with individual constraints subject to privacy, %
and a central operator whose objective is to  %
 optimizing a global, possibly non-convex,  cost %
 while satisfying the  agents' constraints.
   We focus on the practical case of the management of energy consumption flexibilities by the operator of a microgrid.
This paper provides a privacy-preserving algorithm that does compute the optimal allocation of resources, avoiding each agent to reveal her private information
(constraints and individual solution profile) neither to the central operator nor to a third party.
Our method
relies on an aggregation procedure: we maintain a global allocation
of resources, and gradually disaggregate this allocation
to enforce the satisfaction of private contraints,
by a protocol involving the generation of polyhedral cuts and secure multiparty computations (SMC). To obtain these cuts, we use an alternate projections
method \`a la Von Neumann, which is implemented locally by each agent,
preserving her privacy needs.
Our theoretical and numerical results show that the method scales well as the number of agents gets large, and thus can be used to solve  the allocation problem in high  dimension, while addressing privacy issues. 
\end{abstract}

\section{INTRODUCTION}

\textbf{Motivation. } Consider an operator of an electricity microgrid optimizing the joint production schedules of renewable and thermal %
power plants in order to satisfy, at each time period, the consumption constraints of its %
consumers. %
To optimize the costs and the  renewables integration, this operator relies on demand response techniques, that is, taking advantage of the flexibilities of some of the consumers electric appliances---those which can be controlled without impacting the consumer's confort, as electric vehicles or water heaters \cite{PaulinTSG17}. %
 However, for privacy reasons, consumers are not willing to provide   neither their consumption constraints nor their consumption profiles to a central operator or any third party, as this information could be used to induce private information such as their presence at home.

The \emph{global problem} of the operator is to find an allocation  of power (aggregate consumption) $\pp=(p_t)_t$  at each time period (\textit{resource}) $t\in\T$, such that $\pp \in \P$ (feasibility constraints of power allocation, induced by the power plants constraints). Besides, this aggregate allocation has to match an individual comsumption profile $\xx_n= (x\nt)_{t\in\T} $ for each of the consumer (agent) $n\in\N$ considered. The problem can be written as follows:
\begin{subequations}
\label{pb:global}
\begin{align} 
& \min_{\xx\in \rr^{N \times T}\hh,\  \pp\in \P} f(\pp)  \\
 & \xx_n \in \X_n, \   \forall n \in \N   \label{cons:indfeas}\\
 &  \sum_{n\in\N} x\nt = p_t, \ \forall t \in \T \label{cons:disagfeas} \  ,
\end{align}
\end{subequations}
The (aggregate) allocation $\pp$ can be made \emph{public}, that is, revealed to all agents. However, the individual constraint set $\X_n$ and individual profiles $\xx_n$ constitute \emph{private} information of agent $n$, and should not be reavealed to the operator or any third party. %

The approach adopted in our paper is to deal with the problem \eqref{pb:global}
 as two kinds of interdependent subproblems. The firsts are optimal resource allocation problems, or \emph{master problems}, $\min_{\pp\in \P^{(s)}} f(\pp)$ which consists in finding an \emph{aggregate} allocation  over $T$ resources ($\P^{(s)} \subset \P \subset \rr^T)$. %
The second kind is problems of finding the \textit{disaggregation} of a given aggregate allocation $\pp$, that is, to find an individual profile $\xx_n$ for each agent (consumer) $n$ satisfying her individual constraint \eqref{cons:indfeas}, such that the aggregate of the profiles is the optimal allocation \eqref{cons:disagfeas} determined in a master problem.

Aside from the example above, ressource allocation problems (optimizing common resources shared by multiple agents) find many applications in energy \cite{muller2017aggregation,PaulinTSG17}, %
 logistics \cite{laiLam95shipping}, distributed computing \cite{ma1982task}, health care \cite{rais2011operations} and telecommunications  \cite{zulhasnine2010efficient}. 
 In these applications, several entities or agents (e.g. consumers, stores, tasks) share a common resource (energy, products, CPU time, broadband) which has a global cost for the system. %
 For large systems composed of multiple agents, the dimension of the overall problem can be prohibitive and one can rely on decomposition and distributed approaches \cite{ bertsekas1989parallel,palomar2006tutorial,xiao2006optimal}  to answer to this issue.
Besides, %
agents' individual constraints are often subject to privacy issues  \cite{huberman2005valuating}. %
These considerations have paved the way to the development of privacy-preserving, or non-intrusive methods and algorithms, e.g.  \cite{zoha2012non,jagannathan2006new}.

In this work, we consider that each agent has a global demand constraint (e.g. energy demand or product quantity), which confers to the disaggregation problem the particular structure of a transportation polytope \cite{bolker1972transportation}: the sum over the agents is fixed by the aggregate solution $\pp$, while the sum over the $T$ resources are fixed by the agent global demand constraint. Besides, individual constraints can also include minimal and maximal levels on each resource, as for instance electricity consumers require, through their appliances, a minimal and maximal power at each time period.

\newcommand{\wP}{\widetilde{P}}
\textbf{Main Results.} The major contribution of the paper is to provide a non-intrusive and distributed algorithm (\Cref{algo:confidOptimDisag}) that computes an aggregated resource allocation $\pp$,  optimal solution of the---possibly nonconvex---optimization problem \eqref{pb:global}, along with feasible individual profiles $\xx$ for agents, without revealing the individual constraints of each agent to a third party, either another agent or a central operator. 
The algorithm solves iteratively intances of master  problem $\min_{\pp \in\P^{(s)}} f(\pp) $ %
by constructing successive approximations $\P^{(s)} \subset \P$ of the aggregate feasible set of \eqref{pb:global} for which a disaggregation exists, %
by  adding a new constraint on $\pp$  to $\P^{(s)}$, before solving the next master problem. 

To identify whether or not disaggregation is feasible and to add a new constraint in the latter case, our algorithm relies on the alternating projections method (APM) \cite{von1950functional,GUBIN19671} for finding a point in the intersection of convex sets. Here, we consider the two following sets: on the one hand, the affine space defined by the aggregation to a given resource profile, and on the other hand, the set defined by all agents individual constraints (demands and bounds). As the latter is defined as a Cartesian product of each agent's feasibility set, APM can operate in a distributed fashion. The sequence constructed by the APM converges to a single point if the intersection of the convex sets is nonempty, and it converges to a periodic orbit of length $2$ otherwise.
Our key result is the following: if the APM converges to a periodic orbit, meaning that the disaggregation is not feasible,  we construct from this orbit a
polyhedral {\em cut},
i.e. a linear inequality satisfied by all feasible solutions $\pp$  of the global 
problem \eqref{pb:global}, but violated from the current resource allocation (\Cref{thm:violatedCut}). Adding this cut to the master problem, we can recompute a new resource allocation and repeat this procedure until disaggregation is possible. 
  Another major result stated in this paper is the explicit  upper bound on the convergence speed of APM in our framework (\Cref{thm:cvgAP}), which is
  obtained by spectral graph theory methods, exploiting also geometric properties of transportation polytopes.  %
  This explicit speed shows a linear impact of the number of agents, which is a strong argument for the applicability of the method in large distributed systems.

\textbf{Related Work. } A standard approach to solve resource allocation problems in a distributed way is to use a Lagrangian (dual) decomposition technique \cite{palomar2006tutorial, xiao2004simultaneous,seong2006optimal}.
\todo[inline,disable]{SG: I rewrote lagrange *decomposition* and not *relaxation*. An important point is whether any earlier paper has implemented Lagrange decomposition in a privacy-preserving way, using the same kind of SMC technique. If so, we should be more specific. If this has not been done explicitly, it would be better to say something like:
  ``Lagrange decomposition is generally used to reduce the dimension. It may also be implemented in a way which preserve privacy, in a distributed way (see Remark XXX in Section~IV -- add a remark there with a
  comparison between Lagrange relaxation and our method.). However, Lagrange decomposition methods are based on strong duality property, requiring global convexity hypotheses which are not satisfied in many practical problems: for instance, in the field of energy, the master allocation problem is generally a mixed integer linear program (MILP), see \Cref{sec:appli}.  }
  Those techniques are generally used to decompose a large problems into several subproblems of small dimension. They may also be implemented in a way which preserve privacy (see \Cref{rem:lagrangian_privacy} in \Cref{sec:confComputationResources}). However, Lagrangian decomposition methods are based on strong duality property, requiring global convexity hypothesis which are not satisfied in many practical problems (e.g. MILP, see \Cref{sec:appli}).
On the contrary, our method can be used when the master allocation problem is not convex.
In \cite{muller2017aggregation}, the authors study a disaggregation problem similar to the one considered in this paper. Their results concern \emph{zonotopic} sets,  which is different from the structure we described in \Cref{sec:problemDescription}.
The APM has been the subject of several works in itself \cite{GUBIN19671,bauschke1993convergence,bauschke2015bregman}. The authors of \cite{borwein2014analysis} provide general
results on the convergence rate of APM for semi-algebraic sets. They show that the convergence is geometric for polyhedra. However, it is generally hard to compute explicitly the geometric convergence rate of APM, as this requires to bound the singular values of certain matrices arising from the polyhedral constraints. In  \cite{nishihara2014convergence}, the authors provide an explicit convergence rate for APM on a class of polyhedra arising in submodular optimization. The sets they consider differ from the present transportation polytopes.
\textbf{Structure. } In \Cref{sec:problemDescription}, we describe the master resource allocation problem and formulate the associated disaggregation problem. In \Cref{sec:APM}, we focus on the APM and state our main results. In \Cref{sec:confComputationResources}, we apply these results to describe a non-intrusive version of APM (NI-APM) that is used to describe our non-intrusive algorithm for computing an optimal resource allocation. Finally, in \Cref{sec:appli}, we provide a  concrete numerical example based on a MILP to model the management of a  local electricity system (microgrid), and study numerically the influence of the number of agents on the time needed for convergence of our algorithm.

\textbf{Notation.}
Vectors and matrices are denoted by bold fonts, $\bm{v}^\tr$ denotes the transpose of $\bm{v}$, $\dsone_{K}$ denotes the vector $(1 \dots  1)^\tr$ of size $K$, $\U([a,b])$ stands for the uniform distribution on $[a,b]$.
We use $\norm{\xx}_2$ to denote the Frobenius norm $\norm{\xx}^2= \sum_{n,t} x\nt^2$, 
and  $P_C(.)$ to denote the Euclidean projection on a convex set $C$. 
\section{Master problem and disaggregation structure}
\label{sec:problemDescription}

\todo[inline,disable]{SG: I rewrote the next para, this has still to be improved, these sentences are decisive in the reading}
In this work, we suppose an operator
wishes to determine an allocation of resources,
represented
by a $T$-dimensional vector $\pp$, in order to minimize
a global cost function $f$, for instance, an electricity power economic dispatch (or the allocation of different types of merchandise in warehouses in logistics applications) subject to a set of constraints described by a feasibility set $\P$. This problem can be nonconvex either because of nonconvex costs $f$ or because of a nonconvex feasible set $\P$ (see \Cref{sec:appli}). In the proposed method, the operator will consider \emph{master problems} of the form:
\vspace{-0.1cm}
\begin{subequations}
\label{eq:MP}
\begin{align} 
&\min_{\pp\in \rr^T} f(\pp)  \\
 \text{ s.t. } & \pp \in \P^{(s)} \ ,
\vspace{-0.3cm}
\end{align}
\end{subequations}
where the set $\P^{(s)}\subset \P$  is an aggregate approximation  of disaggregation constraints. 
Indeed, the resource allocation $\pp$ has to be shared between $N$ agents (e.g. consumers). Each agent has a global demand (total energy needed) $E_n$ and some lower and upper bounds on each of the  resource $t\in\T$.
The admissible set of profiles of agent $n$ is therefore:
\begin{equation} \label{eq:Xdef}
\hh \X_n \hh\eqDef \hh \{ \xx_n \hh \in\hh \rr^T  | \ \xx_n^\tr \dsoneT \hh = \hh E_n \text{ and }  \forall t, \lbx\nt \hh \leq \hh \x\nt \hh\leq \hh\ubx\nt\}.
\end{equation}
The \textit{disaggregation problem} consists in finding individual profiles $ \xx=(x\nt)\nt \in \rr^{\Ni\Ti}$ of a given aggregated allocation $\pp$ such that $\xx_n$ is feasible for each agent $n$: %
\begin{align} \label{pb:disag}
& \textsc{Find } \xx \in \Y_{\pp} \cap \X  \\
 \text{ where } & \Y_{\pp} \eqd \{\yy\in\rr^{NT} | \yy^\tr \dsoneN  = \pp \} \text{ and } \X \eqd \prod_{n\in\N} \X_n \ . \nonumber
\end{align}
 Following  \eqref{pb:disag}, the \textit{disaggregated} profile refers to $\xx$, while the \textit{aggregated} profile refers to the allocation $\pp$.
 
\begin{figure}[htb]
\begin{center}
\begin{tikzpicture}[scale=0.7]
\node [draw,red,dashed] (h1) at (0,2) {$p_1=0$};
\node [draw] (h2) at (0,0) {$p_2=3$};

\node [draw,red,dashed] (i1) at (5,2) {$E_1=2$};
\node [draw] (i2) at (5,1) {$E_2=0.5$};
\node [draw] (i3) at (5,0) {$E_3=0.5$};

\draw [->] (h1) --  node [text width=2.5cm,midway,above,align=left ] {} (i1);
\draw [->] (h1) --  node [text width=2.5cm,midway,above,align=left ] {} (i2);
\draw [->] (h1) --  node [text width=2.5cm,midway,above,align=left ] {} (i3);
\draw [->,red] (h2) --  node [text width=2.5cm,midway,below,align=left ] {} (i1);
\draw [->] (h2) --  node [text width=2.5cm,midway,below,align=left ] {} (i2);
\draw [->] (h2) --  node [text width=2.5cm,midway,below,align=left ] {} (i3);

\end{tikzpicture}
\end{center}
\caption{Example of disaggregation structure $(T=2,N=3)$, with $\bm{\lbx} =\bm{0}$ and $\bm{\ubx}:=\bm{1}$. \textit{Although the aggregate constraints \eqref{eq:agg-cond} are satisfied, the disaggregation \eqref{pb:disag} of $\pp$ is not feasible in this example (see \Cref{thm:hoffman}).}}
\label{fig:example-disag}
\end{figure}
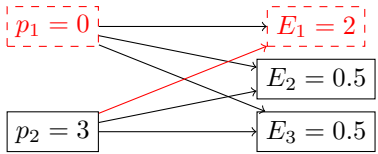

Problem \eqref{pb:disag} may not always be feasible. 
Some necessary conditions for a disaggreagation to exist, obtained by summing the individual constraints on $\N$, are the following \textit{aggregated } constraints:
 \label{eq:agg-cond}
\begin{align}
& \pp^\tr \dsoneT= \EE^\tr \dsoneN \text{ and } \ 
 \llbx^\tr \dsoneN \leq \pp \leq \uubx^\tr  \dsoneN \ .
\end{align}
However, \eqref{eq:agg-cond} are not sufficient conditions, as shown in \Cref{fig:example-disag} where the problem \eqref{pb:disag} is represented as a \textit{flow}  or \textit{circulation} problem from source nodes $t \in \T$ to sink nodes $n \in \N$.

Indeed, with this circulation representation of the disaggregation problem \eqref{pb:disag}, an immediate consequence of Hoffmann theorem \cite[Thm. 3.18]{cook2009combinatorial}\cite{hoffman1960some} is the following characterization of the disaggregation feasibility, which involves an exponential number of inequalities:
\begin{theorem} \label{thm:hoffman}
The disaggregation problem \eqref{pb:disag} is feasible (i.e. $\X \cap \Y_{\pp} \neq \emptyset$) iff for any $\T\mrin  \subset \T, \N\mrin \subset  \N $:
\begin{equation} \label{eq:hoffman}
\sum_{t \notin \T\mrin}\hh  p_t \leq  \sum_{t \notin \T\mrin, n \in \N\mrin} \hh\hh \ol{x}_{n,t}  - \hh\hh \sum_{t \in \T\mrin, n \notin \N\mrin}\hh\hh \ul{x}_{n,t} + \sum_{n \notin \N\mrin} \hh E_n .
\end{equation}
\end{theorem}
The inequality \eqref{eq:hoffman} has a simple interpretation: the residual demand (the left hand side composed of demand and exports minus production) in $\T\mrin \cup \N\mrin$ cannot exceed the import capacity (right hand side of the inequality). One can see that, in the example of \Cref{fig:example-disag}, inequality \eqref{eq:hoffman} does not hold when using the cut composed of the dashed nodes $p_1$ and $E_1$.

There are two main reasons for which solving \eqref{pb:global} is harder  than solving \eqref{eq:MP} and \eqref{pb:disag} separately:
\begin{enumerate}[wide,label=\roman*)]
\item the dimension of \eqref{pb:global} can be huge, as the number of agents $N$ can be really important, for instance in the example of individual consumers;
\item also, and this is the main motivation of this work, the information related to $(\uubx_n)_n, (\llbx_n)_n$ and $(E_n)_n$ might not be available to the centralized operator in charge of optimizing resources $\pp$, as this information may be confidential and kept by each agent $n$, not willing to reveal it to any third party.
\end{enumerate}
In the next sections, we provide a method that addresses those two issues, by considering subproblems \eqref{eq:MP} and \eqref{pb:disag} independently and iteratively, and exploiting the decomposable structure of problem \eqref{pb:disag}.

\section{Alternate Projection Method (APM)} 
\label{sec:APM}
\subsection{Convergence of APM on Transportation Polytopes}

In this section, we consider a fixed aggregated profile $\pp$ and present the Von Neumann Alternate Projections Method (APM) \cite{von1950functional} which solves the problem \Cref{pb:disag} of finding a point in the intersection $\X \cap \Y_{\pp}$.
In the remaining, we will often ommit $\pp$ and just write $\Y$ to denote $\Y_{\pp}$.
The key idea of the method proposed in this paper is to use results of APM   
 to generate a cut in the form of \eqref{eq:hoffman} and to add it as a new constraint in the master problem \eqref{eq:MP} to ``improve'' the aggregated profile $\pp$ for the next iteration.  
As described in \Cref{algo:vonNeumannProj}, APM can be  used to decompose \eqref{pb:disag} and only involves \textit{local} operations. %

\newcommand{\epscvg}{\varepsilon_{\text{cvg}}}
\newcommand{\epsdis}{\varepsilon_{\text{dis}}}
\begin{algorithm}
\begin{algorithmic}[1]
\Require Start with $\yy^{(0)}$, $k=0$ , $\epscvg$, 	 a norm $\norm{.}$ on $\rr^{NT}$ \;
\Repeat
 \State  $\xx^{(k+1)} \leftarrow P_{\X}( \yy^{(k)})$ \;
  \State  $\yy^{(k +1)} \leftarrow P_{\Y}(\xx^{(k+1)})$ \;
\State  $k \leftarrow k+1$ \;
\Until{$\norm{\yy^{(k)} -\yy^{(k-1)} } < \epscvg $}
\end{algorithmic}
\caption{Alternate Projections Method (APM)}
\label{algo:vonNeumannProj}
\end{algorithm}
The convergence of \Cref{algo:vonNeumannProj} is proved by \Cref{thm:cvgAP}:
\begin{theorem}[\cite{GUBIN19671}] \label{thm:cvgAP}Let $\X$ and $\Y$ be two convex sets with $\X$ bounded, and let  $(\xE\kexp)_k $ and $(\xP\kexp)_k$ be the two infinite sequences generated by \Cref{algo:vonNeumannProj} with $\epscvg=0$. Then there exists $\xx^\infty \in \X$ and $\yy^\infty\in \Y $ such that:
\begin{subequations}
\begin{align}
& \xE\kexp \underset{k\rightarrow \infty}{\longrightarrow} \xx^\infty \ , \quad \xP\kexp \underset{k\rightarrow \infty}{\longrightarrow} \yy^\infty ; \\
& \norm{\xx^\infty - \yy^\infty }_2= \min_{{\xx \in \X , \yy \in \Y } } \norm{ \xE - \xP}_2 \ . 
\end{align}
\end{subequations}
In particular,  if  $\X \cap \Y \neq \emptyset$, then $(\xE\kexp)_k $ and $(\xP\kexp)_k$ converge to a same point $\xx^\infty \in \X \cap \Y$.

\end{theorem}
If disaggregation is not feasible, \Cref{thm:cvgAP} states that  APM will ``converge'' to an orbit $(\xx^\infty ,\yy^\infty)$ of period 2.

The convergence rate of APM has been the subject of several works \cite{bauschke1993convergence,borwein2014analysis}, and it strongly depends on the structure of the sets on which the projections are done: for instance, if the sets are polyhedral, \cite[Prop.~4.2]{borwein2014analysis} shows that the convergence is geometric. However, there are very few cases in which an explicit upper bound on the convergence rate has been proved. In our case, we are able to obtain such a bound, as shown in the following theorem:
\begin{theorem} \label{th:boundCvgRate} 
For the  sets $\X$ and $\Y$ defined in (\ref{eq:Xdef}-\ref{pb:disag}), the two subsequences of alternate projections converge at a geometric rate to $\xEinf\in \X$, $\xPinf \in \Y$, with: 
\begin{align*}
&\hh \| \xx^{(k)} \hh -\xEinf \|_2 \hh \leq\hh  2 \| \xx^{(0)} \hh -\xEinf \|_2 \times \rhoN^k \\
\text{ where } & \rhoN \eqd   1- \tfrac{1}{4}\left( {N(T+1)^2(T-1)} \right) <1 \ , 
\end{align*}
Same inequalities hold for the convergence of $\yy^{(k)}$ to $\xPinf$.
\end{theorem}
\proof \Cref{app:proofCvgRate} provides a sketch of the proof. %

\todo[inline,disable]{SG: In the appendix (see my remark above), include some key elements of proof}

\Cref{th:boundCvgRate} shows that the APM is efficient in our case of bounded transport polytopes. It shows that the number of iterations for a given accuracy grows linearly in the number of agents $N$. %

As stated in \eqref{pb:disag}, the set $\X$ is a Cartesian product $\prod_n \X_n$, so that the projection \eqref{pb:QPxE} can be computed by $N$ projections on $(\X_n)_n$, which can be executed in parallel. Now, instead of solving the quadratic program by standard interior point methods and due to its particular structure, we can use the algorithm of Brucker \cite{brucker1984n}, which has a complexity in $\mathcal{O}(T)$.
On the other hand,  $P_\Y(.)$ is a projection on an affine space, 
 and the solution  can be obtained explicitly as:
\begin{equation} \label{eq:explicitxP}
 \forall n,t, \xP\ih= \xE\ih + \nu_\h \text{ and } \nnu = \tfrac{1}{N} (\pp-  \xE^\tr \dsoneN) \ .
\end{equation}

\subsection{Generation of a cut from APM iterates}
Our key result is the following: in the case where APM converges to a periodic orbit $(\xx^\infty ,\yy^\infty)$  with $\xEinf \neq \xPinf$ (see \Cref{thm:cvgAP}), we obtain
 from $(\xx^\infty ,\yy^\infty)$ an inequality \eqref{eq:hoffman} that is violated by $\pp$:
\begin{theorem} \label{thm:violatedCut}
For the  sets $\X$ and $\Y$ defined in (\ref{eq:Xdef}-\ref{pb:disag}) and if  $ \X \cap \Y = \emptyset$ %
, the following sets given by the limit orbit $(\xx^\infty ,\yy^\infty)$ defined in \Cref{thm:cvgAP}:
\begin{subequations}
\label{eq:H0N0withoutbounds}
\begin{align} 
& \T_0 \eqDef \{ t | p_t >  \txt\sum_{n\in\N} x^\infty\nt \} \\ 
& \N_0 \eqDef \{ n \ | E_n - \txt\sum_{t\notin \T_0} \lbx\ih - \sum_{t\in\T_0} \ubx\ih <0 \} \   
\end{align}
\end{subequations}
define a Hoffman cut of form \eqref{eq:hoffman} violated by $\pp$, that is:
\begin{equation}\label{eq:cutViolatedH0N0asym}
\sum_{n\in\N_0} E_n - \sum_{t\in\T_0} p_t + \sum_{t \in \T_0, n \notin \N_0} \hh\hh\hh\ubx\ih  - \sum_{t \notin \T_0, n \in \N_0} \hh\hh\hh \lbx\ih  < 0 \ .
\end{equation}
This cut can be reformulated in terms of $\dsoneN^\tr \xEinf$ as:
\begin{equation}  \label{eq:cutConfidWithoutN0}
\hh \A_{\T_0}  \hh < \hh \sum_{t\in\T_0} \hh  p_t \text{\  with }  \A_{\T_0}\eqDef \hh \sum_{t\in\T_0} \sum_{n\in\N} \hh \hh \x^\infty\nt  .
\end{equation}
\end{theorem}
\proof \Cref{app:proofClaimFacts} gives the sketch of the proof of \Cref{thm:violatedCut}. The complete proofs will be given elsewhere.

One can see that, intuitively, $\N_0$ is the subset associated to $\T_0$ that minimizes the right hand side of \eqref{eq:hoffman}.
Note that \Cref{thm:violatedCut} gives an alternative constructive proof  of Hoffman circulation's theorem (\Cref{thm:hoffman}) in the case of a bipartite graph of the form of \Cref{fig:example-disag}. 
Moreover, in the case where the disaggregation problem \eqref{pb:disag} is not feasible, the negation of equation \eqref{eq:cutConfidWithoutN0} provides a new valid constraint  as a condition for the existence of a  disaggreagated profile of $\pp$. This constraint can be used in the master problem  \eqref{eq:MP} to update the vector of resources $\pp$ for the next iteration. %
 This constraint only involves the \textit{aggregate} information  $\dsoneN^\tr \xEinf$ on the users profile. To make the process fully \emph{non-intrusive}, we explain in \Cref{subsec:NIAPM} how the operator can compute this constraint without making the agents reveal their profiles $(\xEinf_n)_{n\in\N}$.

\section{Non-intrusive Projections and Computation of Disaggregated Optimal Resources}
\label{sec:confComputationResources}
\newcommand{\dsg}{ \textsc{Disag} }
\newcommand{\ids}{ ^{(s)} }
\newcommand{\cs}{_\text{cs}}
\newcommand{\ssig}{\bm{\sigma}}

\subsection{Non-Intrusive Alternate Projections Method (NI-APM)} \label{subsec:NIAPM}
Because of the particular structure of the problem, the projections in APM  can be computed  separately by the operator and the agents. The projection $P_\Y$ is made by the operator, which only requires to know $\pp$ and the aggregate profile $\xE^\tr \dsoneN$ according to from \eqref{eq:explicitxP}.  The projection $P_\X$ on $\X= \prod_n \X_n$ is executed in parallel by each agent: $n$ computes $P_{\X_n}$ which only needs her private information $E_n$ and $\uubx_n,\llbx_n$.
 However, in the way APM is described in \Cref{algo:vonNeumannProj}, the operator and the agents still need to exchange the iterates $\xE^{(k)},\xP^{(k)}$ at each step. 
To avoid the transmission of agents' profiles %
 to the operator, we use a secure multiparty computation (SMC) technique (see \cite{yao1986secrets}) which enables the operator to obtain the aggregate profile $\SS^{(k)}:=  \dsoneN^\tr \xE^{(k)}$ in a non-intrusive manner, as described in \Cref{algo:SMCsum}.

The main idea of SMC is that, instead of sending her profile $\xE_n$,  agent  $n$ \emph{splits} $\x\nt$  for each $t$ into $N$ random parts $(s_{n,t,m})_m$, according to an uniform distribution and summing to $\xE\nt$ (Lines \ref{algline:SMC:cut1}-\ref{algline:SMC:cut2}). Thus, each part $s_{n,t,m}$ taken individually does not reveal any information on $ \xE_n$ nor on $\X_n$, and can be sent to agent $m$. 
Once all exchanges of parts are completed (Line 5), and $n$ has herself received the parts from other agents, agent $n$ computes a new aggregate quantity $\ssig_n$ (Line \ref{algline:NIAPM-computesigma}), which does not contain either any information about any of the agents, and sends it to the operator (Line \ref{algline:NIAPM-Sendsigma}). The operator can finally compute the quantity $\SS=\xE^\tr \dsoneN= \ssig^\tr \dsoneN$. 

\begin{algorithm}[!ht]
\begin{algorithmic}[1]
\Require Each agent has a profile $(\xx_n)_{n\in \N} $\;
	\For{each agent $\i \in \N$}
  \State Draw $\forall t, (s_{n,t,m})_{m=1}^{N-1} \hh \in\hh \mathcal{U}([0,A]^{N-1})$ \; \label{algline:SMC:cut1}
  \State and set $\forall t,  s_{n,t,N}\hh \eqd \hh {\x}_{n,t}- \sum_{m=1}^{N-1} s_{n,t,m}$\; \label{algline:SMC:cut2}
  \State Send $(s_{n,t,m})_{t\in\T} $ to agent $m \in \N$ \;
  \EndFor
  \For{each agent $\i \in \N$}
  \State Compute $\forall t, \sigma\ih = \sum_{m\in\N} s_{m,t,n} $  \label{algline:NIAPM-computesigma}\;
  \State Send $(\sigma\ih)_{t\in\T}$ to operator \label{algline:NIAPM-Sendsigma}\;
  \EndFor \label{algline:NIAPM-SendsigmaOp} 
  \State Operator computes $\SS= \sum_{n\in\N} \ssig_n$ \;
\end{algorithmic}
\caption{SMC of Aggregate (SMCA) $\sum_ {n\in\N} \xx_n$}
\label{algo:SMCsum}
\end{algorithm}
\begin{remark} As  $\bm{\sigma}_n,$ and $\ss_n$ are random by construction, an eavesdropper aiming to learn the profile $\xx_n$ of $n$ has no choice but to intercept all the communications of $n$ to all other agents (to learn $(s_{n,t,m})_{m\neq n}$ and $(s_{m,t,n})_{m\neq n}$) and to  the operator (to learn $\sigma_n$). To increase the confidentiality of the procedure, one could use any  encryption scheme (such as RSA \cite{rivest1978method}) for all  communications involved in \Cref{algo:SMCsum}.
\end{remark}

We can use this non-intrusive computation of aggregate $\SS$ in APM to obtain a \emph{non-intrusive} algorithm \textit{NI-APM} (\Cref{algo:vonNeumannProjasymConfid}) in which agents do not reveal neither their profiles nor their constraints to the operator.
\renewcommand{\B}{B} %
\begin{algorithm}[!ht]
\begin{algorithmic}[1]
\Require Start with $\yy^{(0)}$, $k\hh=\hh0$, $\epscvg,\epsdis$, norm $\norm{.}$ on $\rr^{NT}$ \;
\Repeat \label{algline:NIAPM-mainLoop}
 	\For{each agent $\i \in \N$}
  \State  ${\xx}_n^{(k)} \leftarrow P_{\X_n}( {\yy}_n^{(k-1)})$  \label{algline:NIAPM-compute-x}\;
  \EndFor
 \State Operator obtains $\SS^{(k)} \leftarrow $SMCA($\xx^{(k)}$)  (\textit{cf} Algo.\ref{algo:SMCsum}) \label{algline:NIAPM-optain-agg}\;
\State  and sends $\nnu^{(k)}\hh:=\frac{1}{N}( \pp-\SS^{(k)} ) \in \rr^T$ to agents $\N$ \label{algline:NIAPM-opsendsnu}\;
  	\For{each agent $\i \in \N$}
  	\State Compute $\xP_\i^{(k)} \leftarrow \xE_\i^{(k)} + \nnu^{(k)} $ \label{algline:NIAPM-compute-y}\;
\EndFor
\State  $k \leftarrow k+1$ \;
\Until{ $ \norm{\xx^{(k)} - \xx^{(k-1)}  } < \epscvg$ } \label{algline:NIAPM-conv-APM}
\If { $\norm{\xx^{(k)} - \yy^{(k)} } \leq \epsdis$} \label{algline:NIAPM-conv-disag}

\hspace{-1cm}\Lcomment{found a $\epsdis$-solution of the disaggregation problem}
\State  Each agent adopts profile $\xE_\i^{(k)} $ \;	
\State \Return \textsc{Disag} $\leftarrow$ \textsc{True} \;
\Else 
\Lcomment{have to find a valid constraint violated by $\pp$} 
\State Operator computes $\T_0 \leftarrow \{t \in \T \ | \ \tfrac{3}{2} B\epscvg < \nu_t^{(k)}  \}$  \label{algline:NIAPM-defH0asymConfid}\;
\State Operator computes $\A_{\T_0} \eqd \sum_{\h\in\T_0} 	\dsoneN^\tr \ssig_t^{(k)}  $   \;
\If {$\A_{\T_0}- \sum_{t\in\T_0} \pp_t <0$} \label{algline:NIAPM:checkconstraint}
\State \Return \textsc{Disag} $\leftarrow$ \textsc{False}, $\A_{\T_0}$\; 
\Else 
\Lcomment{need to run APM with higher precision}
\State Return to Line~\ref{algline:NIAPM-mainLoop} with $\epscvg \leftarrow \epscvg/2$ \label{algline:NIAPM-resolve}\;
\EndIf \;
\EndIf \;
\end{algorithmic}
\caption{Non-intrusive APM %
 (NI-APM)}
\label{algo:vonNeumannProjasymConfid}
\end{algorithm}

One can see that $\xE$ and $\xP$ computed in Lines \ref{algline:NIAPM-compute-x} and \ref{algline:NIAPM-compute-y} in \Cref{algo:vonNeumannProjasymConfid} correspond to the projections computed in the original APM \Cref{algo:vonNeumannProj}. 
In \Cref{algo:vonNeumannProjasymConfid},  the operator obtains the aggregate profile $\SS^{(k)}$ (Line \ref{algline:NIAPM-optain-agg}),  computes and sends the corrections $\nnu^{(k)}$ to all agents (Line \ref{algline:NIAPM-opsendsnu}). Then, each agent can compute locally the projection $\xP_\i^{(k)} =P_{\Y}(\xE_\i^{(k)})$ by applying the correction  $\nnu^{(k)}$ (Line \ref{algline:NIAPM-compute-y}). %

Using \eqref{eq:explicitxP}, we get $\nnu^{(k)} \rightarrow \nnu^{\infty} \eqd \tfrac{1}{N} (\pp-  \dsoneN^\tr\xE^\infty)$.  
 \Cref{thm:violatedCut} %
 uses this limit value through %
 ${\T_0}^\infty \eqd \{t \in \T  |  0 < \nu_t^{\infty}  \}$. 
Yet, from APM, one can only access to $\nnu^{(k)}$ and thus to the \emph{approximation} $\T_0$, computed on Line \ref{algline:NIAPM-defH0asymConfid}), where $B$ is a pre-defined constant. %
 However,  we show that for $\epscvg$ small enough and a well-chosen value of $B$, we obtain  $\T_0=\T_0^\infty $, so that  we get the termination result: %
 \newcommand{\rcvg}{r_{\mr{cvg}}}
  \begin{proposition}
  \label{prop:terminationAlgo} For $B> (1-\rhoN)^{-1}$, \Cref{algo:vonNeumannProjasymConfid} terminates in finite time.
  \end{proposition}
  The termination of the loop Lines \ref{algline:NIAPM-mainLoop}-\ref{algline:NIAPM-conv-APM} is ensured by \Cref{th:boundCvgRate}. In the case where $\norm{\xx^{(k)} - \yy^{(k)} } \leq \epsdis$, \Cref{algo:vonNeumannProjasymConfid} terminates. Otherwise, if $\norm{\xEinf-\xPinf}> \epsdis$, then %
 \Cref{algo:vonNeumannProjasymConfid} terminates (i.e. Line \ref{algline:NIAPM:checkconstraint} is True and a new cut is found) as soon as %
 $B \epscvg < \min \left\{ \frac{\norm{\xEinf-\xPinf}_1 }{2\sqrt{N}},\ \frac{2}{5}  \lnu \right\}$, where $\lnu \eqd \min \{|\nu_t^\infty| >0 \} $ and with $\norm{.}=\norm{.}_2$. 
  The complete proof is ommited here. %
  
 In practice, we can start with a large $\epscvg$ to obtain the first constraints while avoiding useless computation, and then half $\epscvg$ if needed (Line \ref{algline:NIAPM-resolve}) until the termination condition holds.%

\begin{remark} \label{rem:lagrangian_privacy}
SMC techniques could also be used to implement non-intrusive Lagrangian decomposition methods. %
 However, these methods rely on a convexity hypothesis that we do not need in the proposed method.
\end{remark}

\subsection{Non-intrusive Disaggregation of Optimal Allocation}

In this section, we describe a method to compute a solution of the global problem \eqref{pb:global}, that is, an optimal resource allocation $\pp$ for which a disaggregation exists, along with an associated disaggregated profile $\xx_n$ for each agent $n$. This computation is done in a non-intrusive manner: the operator in charge of $\pp$ does not have access neither to the bounding constraints $\uubx$ and $\llbx$ of the agents nor to the agents disaggregated profile $\xx$, as detailed in \Cref{algo:confidOptimDisag} below.

\begin{algorithm}[!ht]
\begin{algorithmic}[1]
\Require $s=0$ , $\P^{(0)}=\P$ ;
\dsg = \textsc{False} \;
\While{Not \dsg}
	\State Compute $\pp\ids= \arg\min_{ \pp \in \P\cs\ids } \pp$ \label{algline:Optim:masterProb} \;
	\State \dsg $\leftarrow$ NI-APM($\pp\ids$) \;
	\If{ \dsg }  \label{algline:Optim:disagTrue}
	\State Operator adopts $\pp\ids $ \;
	\Else \label{algline:Optim:disagNot}
 \State Obtain $ \T_0\ids, \A_{\T_0}\ids $ from NI-APM($\pp\ids$) \; 
	\State $\P^{(s+1)} \leftarrow \P\ids \cap \{ \pp | \sum_{t\in\T_0\ids} p_t \leq  \A_{\T_0}\ids \}$ \label{algline:Optim:addConstraint}\;
 \EndIf \;
 \State $s \leftarrow s+1 $\;
\EndWhile  \;
\end{algorithmic}
\caption{Non-intrusive Optimal Disaggregation}
\label{algo:confidOptimDisag}
\end{algorithm}

\Cref{algo:confidOptimDisag} iteratively calls  NI-APM (\Cref{algo:vonNeumannProjasymConfid}) and in case disaggregation is not possible (Line \ref{algline:Optim:disagNot}), a new constraint is added (Line \ref{algline:Optim:addConstraint}), obtained from the quantity $\A_{\T_0}$ defined in \eqref{eq:cutConfidWithoutN0}, to the resource problem \eqref{eq:MP}. This constraint is an inequality on $\pp$ and thus does not reveal significant individual information to the operator. The algorithm stops when disaggregation is possible (Line \ref{algline:Optim:disagTrue}). The termination of \Cref{algo:confidOptimDisag} is ensured by the following property and the form of the constraints added \eqref{eq:cutViolatedH0N0asym}:
\begin{proposition} \label{prop:finiteNumberCuts}
\Cref{algo:confidOptimDisag} stops after a finite number of iterations, as at most $2^T$ constraints (Line \ref{algline:Optim:addConstraint}) can be added to the master problem (Line \ref{algline:Optim:masterProb}).
\end{proposition}
Although there exist some instances with an exponential number of independent constraints, this does not jeopardize the proposed method: %
 in practice, the algorithm stops after a very small number of constraints added (see the example of \Cref{sec:appli}). Intuitively, we will only add constraints ``supporting'' the optimal allocation $\pp$.

Thus, \Cref{algo:confidOptimDisag} is a method which enables the operator to compute a resource allocation $\pp$ and the $N$  agents to adopt profiles $(\xx_n)_n$, such that $(\xx,\pp)$ solves the global problem \eqref{pb:global}, and the method ensures that both:
\begin{enumerate}
\item the information relative to each agent constraints (upper bounds $\uubx_n$, lower bounds $\llbx_n$, demand $E_n$);
\item the final disaggregated profile $\xx_n $ (as well as the iterates $(\xE^{(k)})_k$ and $(\xP^{(k)})_k$ in NI-APM)
\end{enumerate}
are kept confidential by agent $n$ and can not be induced by a third party (either the operator or any other agent $m\neq n$).

\section{Application to Management of a Microgrid}
\label{sec:appli}

We apply the proposed method to solve a nonconvex distributed problem in the energy field. We consider a microgrid \cite{katiraei2008microgrids}  composed of N electricity consumers with flexible appliances (such as  electric vehicles or water heaters), a photovoltaic (PV) power plant and a conventional generator. 
\subsection{Mixed Integer Problem Formulation}
The operator responsible of the microgrid aims at satisfying the demand constraints of consumers over a set of time periods $\T=\{1,\dots,T\}$, while minimizing the energy cost for the community.
We have the following characteristics:
\newcommand{\ppv}{p^{\textsc{pv}}}
\newcommand{\pppv}{\bm{p}^{\textsc{pv}}}

\newcommand{\pg}{p^g}
\newcommand{\ppg}{\bm{p}^g}
\newcommand{\pgub}{\overline{p}^g}
\newcommand{\pglb}{\underline{p}^g}
\newcommand{\I}{\mathcal{I}}
\newcommand{\xst}{\bo^{\textsc{st}}}
\newcommand{\Cst}{C^{\textsc{st}}}
\newcommand{\xon}{\bo^{\textsc{on}}}

\newcommand{\bo}{b} %
\begin{itemize}[wide]
\item the PV plant generates a nondispatchable power profile $(\ppv_t)_{t\in\T}$ at marginal cost zero;
\item the conventional generator has a starting cost $\Cst$, minimal and maximal power production $\pglb,\pgub$, and piecewise-linear and continuous generation cost function $\pg \mapsto f(\pg)$:
\vspace{-0.3cm}
\begin{equation*}
f(\pg)= \alpha_k + c_k \pg   , \text{ if } \pg \in \I_k \eqd [\theta_{k-1}, \ \theta_{k} [, \ k=1\dots K ,
\end{equation*}
where $\theta_0\eqd 0$ and $\theta_K\eqd \pgub$;
\item each agent $n\in\N$ has some flexible appliances which require a global energy demand $E_n$ on $\T$, and has consumption constraints on the total household consumption, on each time period $t\in\T$, that are formulated with $\llbx_n,\uubx_n$. These parameters are confidential because they could for instance contain some information on agent $n$ habits.
\end{itemize}
The master problem \eqref{eq:MP} can be written as a MILP \eqref{modelMG}:
\begin{subequations} \label{modelMG}
\begin{align}
& \hspace{-9pt} \min_{\pp,\ppg, (\ppg_k),(\bm{\bo}_k),\bm{\xon}, \bm{\xst} }  \sum_{t\in\T} \Big( \alpha_1 \xon_t+ \sum_k c_k {\pg_k}_t + \Cst \xst_t \Big) \label{eq:MGobj}\\
\label{eq:formPcwGenCost1} &\pg_t= \txt\sum_{k=1}^K \pg_{k,t} ,  \ \forall t \in \T \\
& \bo_{1,t} \theta_1 \leq  \pg_{1,t} \leq \theta_1 ,  \ \forall t \in \T\\
& \bo_{2,t} (\theta_2- \theta_1) \leq  \pg_{2,t} \leq \bo_{1,t}(\theta_2- \theta_1) ,  \ \forall t \in \T \\
& \vdots \hspace{3cm}\vdots \\
\label{eq:formPcwGenCost2}&  0  \leq  \pg_{K,t} \leq \bo_{K-1,t} (\theta_{K}-\theta_{K-1}) ,  \ \forall t \in \T \\
\label{eq:consStart}& \xst_t \geq  \xon_{t}-\xon_{t-1}, \ \forall t \in \{2,\dots,T\}\\ 
\label{eq:consOn}& \pglb \xon_t \leq \pg_t \leq \pgub \xon_t  ,  \ \forall t \in \T\\
& \xon_t,\xst_t, \bo_{1,t}, \dots, \bo_{K-1,t} \in \{0,1\} ,  \ \forall t \in \T \\
\label{eq:consProdTot}& \pp\leq \pppv + \ppg \\
\label{eq:consProdDemand}& \pp^\tr \dsoneT= \EE^\tr \dsoneN \\
\label{eq:consProdBounds}& \llbx^\tr \dsoneN \leq \pp \leq \uubx^\tr  \dsoneN \ .
\end{align}
\end{subequations}
\todo[inline,disable]{SG: I would give more details. In particular, say that the source of nonconvexity is the presence of the starting cost, controlled by the Boolean variable $x_t^{ST}$. Excepting this cost, which induces a discontinuity, is the fonction $f$ convex? It is also intringuing that only $\alpha_1$, not $\alpha_2,]dots, \alpha_K$ appear in the MILP? Is it due to the continuity? This should be said. The role of the $\bo$ variables should be explained, it is not trivial. Is the MILP valid even if the marginal costs $x_k$ are not increasing? if yes say so.}

In this formulation (\ref{eq:formPcwGenCost1}-\ref{eq:formPcwGenCost2}) are a mixed integer formulation of the generation cost function $f$: one can show that the boolean variable $\bo_{k,t}$ is equal to one \textit{iff} $\pg_{k,t} \geq \theta_k$ for each $k\in \{1,\dots,K-1\}$. Note that only $\alpha_1$ appears in \eqref{eq:MGobj} because of the continuity assumption on $f$.

Constraints (\ref{eq:consStart}-\ref{eq:consOn}) ensure the on/off and starting constraints of the power plant, \eqref{eq:consProdTot} ensures that the power allocated to   consumption is not above the total production, and (\ref{eq:consProdDemand}-\ref{eq:consProdBounds}) are the aggregated feasibility conditions already referred to in \eqref{eq:agg-cond}. Note that more complex and realistic MILP models exist for power plants (e.g. \cite{carrion2006computationally}), but with the same structure than \eqref{modelMG}.
The nonconvexity of \eqref{modelMG} comes from the existence of starting costs and on minimal the power constraint, which makes necessary to use boolean state variables $\xst,\xon$.

\subsection{Parameters}
\begin{figure}[ht]
\centering
\includegraphics[width=0.9\columnwidth]{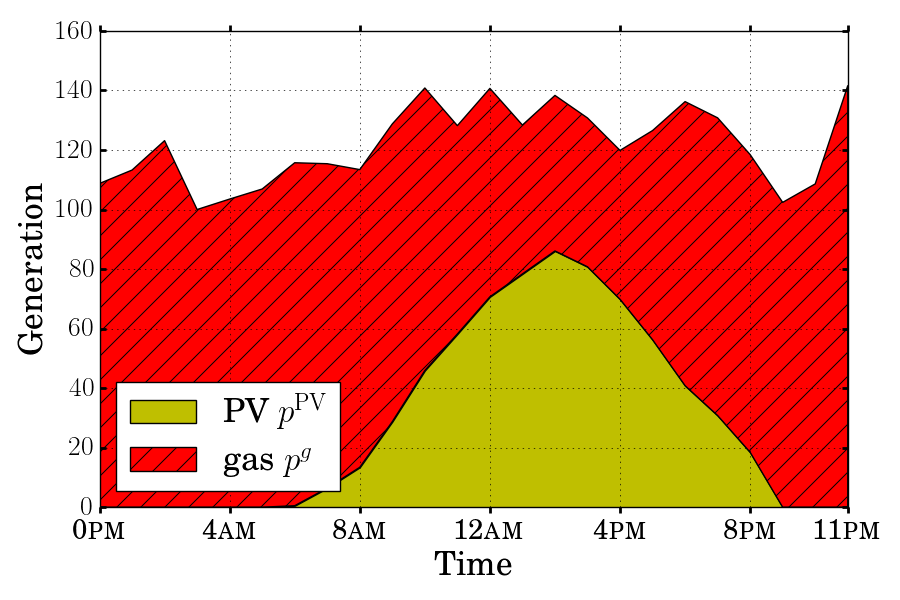}
\caption{Example of optimal resource allocation (power production $\pp=\pppv+\ppg$) in the example of \Cref{sec:confComputationResources}, with $N=20$ agents. \textit{Exploiting consumption flexibilities, the consumption is higher during the PV production periods.}}
\label{fig:prod}
\end{figure}

\newcommand{\scal}{\kappa}
We simulate the problem described above for different values of $N\in \{ 2^4$,$2^5$, $2^6$,$2^7$,$2^8 \}$ and one hundred  instances with random parameters for each value of $N$. A scaling factor $\scal_N\eqd N/20$ is applied on parameters to ensure that production capacity  is large enough to meet consumers demand. The parameters are chosen as follows:
\begin{itemize}[wide,labelindent=0pt]
\item $T=24$ (hours of a day);
\item production costs: $K=3$ , $\theta=[0,70,100,300]\scal_N, \bm{c}=[0.2,0.4,0.5]$, $\pglb\hh =\hh50\scal_N , \pgub\hh=\hh300\scal_N$, $\alpha_1 \hh =\hh 4$ and $\Cst=15$;
\item photovoltaic: $\ppv_t\hh\hh=\hh\left[ 50 (  1\hh-\hh\mathrm{cos}(\frac{ (t-6)2\pi}{16})\hh+\hh \U([0,10])\right]\hh\scal_N$ for $t\in \{6,\dots,20\}$, $\ppv_t=0$ otherwise (see \Cref{fig:prod});
\item for consumption parameters, we used $\lbx\nt \sim \U([0,10])$, $\ubx\nt \sim \U([0,5])+\lbx\nt $ and $E_n \sim  \U([ \dsoneT^\tr \llbx_n, \dsoneT^\tr \uubx_n])$, so that individual feasibility ($\X_n \neq \emptyset$) is ensured.%
\end{itemize}
\renewcommand{\arraystretch}{1.2}
\begin{table}[ht]
\centering
\begin{tabular}{|l|c|c|c|c| c|}
\hline 
$N=$ & $2^4$           &  $2^5$ &  $2^6$&  $2^7$ & $2^8$ \\\hline \hline 
\# master pb. &  193.6	   &  194.1 & 225.5 &  210.9  & 194.0  \\\hline 
\# projs.& 9506.9      & 15366.7 &24319.3& 26537.5& 26646.4  \\\hline%
\end{tabular}
\caption{number of subproblems solved  (average on 100 instances)}%
\label{tab:NumberSubpb}
\end{table}

\begin{figure}[ht]
\vspace{-0.8cm}
\centering
\includegraphics[width=0.75\columnwidth]{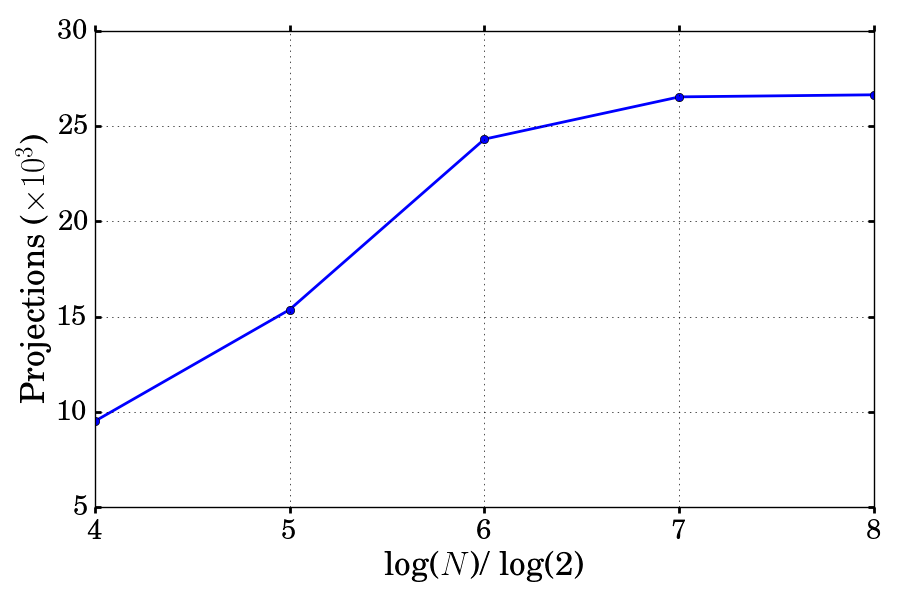}
\vspace{-0.3cm}
\caption{Total number of computed projections for different values of $N$. \textit{We observe that the number of agents $N$ has a sublinear impact on the total number of projections needed. %
}}
\label{fig:numberProjs}
\end{figure}
\subsection{A limited impact of the number $N$ of agents}
We implement \Cref{algo:confidOptimDisag} using Python 3.5. The MILP  \eqref{modelMG} is solved using Cplex Studio 12.6 and  Pyomo interface. Simulations are run on a single core  of a cluster at 3GHz. %
For the convergence criteria (see Lines \ref{algline:NIAPM-conv-APM} and \ref{algline:NIAPM-conv-disag} of \Cref{algo:vonNeumannProjasymConfid}), we use $\epsdis=0.01$ with the operator norm  defined by $\normop{\xx}= \max_{n\in \N} \sum_t |x\nt|$ (to avoid the $\sqrt{N}$ factor in the convergence criteria appearing with $\norm{.}_2$), and starts with $\epscvg=0.1$. 
The largest instances took around 10 minutes to be solved  in this configuration and without parallel implementation. As the CPU time needed %
depend on the cluster load, it is not a reliable indicator of the influence on $N$ on the complexity of the problems. 
Moreover, one advantage of the proposed method is that the projections in APM can be computed locally by each agent in parallel, which could not be implemented here for practical reasons.
 
\Cref{tab:NumberSubpb} gives two robust indicators of the influence of $ N$ on the problem complexity: %
the number of master problems solved and the total number of projections computed, 
 on average over the hundred instance  for each value of $N$:
\begin{itemize}[wide]
\item one observes that the number of master problems solved (MILP \eqref{modelMG} ), which corresponds to the number of constraints or ``cuts'' added to the master problem, remains almost constant  when $N$ increases;
\item in all instances, this number is way below the upper bound of $2^{24}> 1,6\times 10^7$ possible constraints (see proof of \Cref{prop:finiteNumberCuts}), which suggests that only a polynomial number of constraints are added in practice;%
\item the average total number of projections computed for each instance (total number of iterations of the \textbf{while} loop of \Cref{algo:vonNeumannProjasymConfid}, Line \ref{algline:NIAPM-mainLoop} over all calls of APM in the instance) increases in a sublinear way, as illustrated in \Cref{fig:numberProjs}, which is even better that one could expect from the upper bound given in \Cref{th:boundCvgRate}.
\end{itemize}
\section{Conclusion} \label{sec:conclusion}
We provided a non-intrusive algorithm that enables to compute an optimal resource allocation, solution of a--possibly nonconvex--optimization problem, and affect to each agent an individual profile satisfying a global demand and lower and upper bounds constraints. 
Our method uses local projections and works in a distributed fashion. Hence, it ensures that the problem is not affected by the high dimension relative to the large number of agents, and that it is privacy-preserving, as agents do not need to reveal any information on their constraints or their individual profile to a third party.
Several extensions and generalizations can be considered.
First, we could generalize the abstract circulation problem on the bipartite graph depicted in \Cref{fig:example-disag} to an arbitrary network, where the  set of nodes is partitioned in $K$ parts defining $K$ sets on which we could make alternating projections.  Second, our method relies on the particular structure obtained from the form of constraints. Although these kind of constraints are widely used to model many practical situations, it would also be useful to obtain similar results for arbitrary (or polyhedral) agents constraints. Last, a deeper  complexity analysis, with a thinner upper bound on the maximal number of constraints (cuts) added in the algorithm (see \Cref{prop:finiteNumberCuts} and \Cref{tab:NumberSubpb}) would constitute interesting results.%
\todo[disable]{SG: the last problem is rather difficult theoretically, perhaps may be rephrased in a less ambitious way (study the complexity)}

\bibliographystyle{IEEEtran}

\bibliography{../../../USEFULPAPERS/Biblio_complete/shortJournalNames,../../../USEFULPAPERS/Biblio_complete/biblio1,../../../USEFULPAPERS/Biblio_complete/biblio2,../../../USEFULPAPERS/Biblio_complete/biblio3,../../../USEFULPAPERS/Biblio_complete/biblioBooks}

\begin{appendices}
  \crefalias{section}{appsec}

\section{Proof of \Cref{prop:factsOncut}}
\label{app:proofClaimFacts}

To show \Cref{thm:violatedCut}, 
 we formulate the projections $P_\X$ and $P_\Y$  as the solutions of the constrained quadratic programs:
\begin{subequations}
\label{pb:QPxE}
\begin{align}
 &\min_{\xx \in \rr^{NT}} \frac{1}{2} \norm{\xx-\yy}_2^2 \\
& \ {\xE}\dsoneT= \EE & ( \bm{\lambda} )\\
&   \ \llbx  \leq {\xE} \leq \uubx   & (\uumu , \oomu)
\end{align}
\end{subequations}
\vspace{-0.5cm}
and:
\vspace{-0.2cm}
\begin{subequations}
\label{pb:QPxP}
\begin{align}
 &\min_{\xP \in \rr^{NT}} \frac{1}{2} \norm{\xx-\yy}_2^2 \\
&   {\yy}^\tr \dsoneN = \pp  & (\nnu) \ ,
\end{align}
\end{subequations}
where $\bm{\lambda}, \uumu, \oomu, \nnu$ are the Lagrangian multipliers associated to the constraints.
Although there is no such explicit characterization of the solution of \eqref{pb:QPxE} as the one \eqref{eq:explicitxP} given for \eqref{pb:QPxP}, we can obtain the following properties:
\begin{proposition}  \label{prop:factsOncut} Suppose that $\X \cap \Y = \emptyset$ and consider the sets $\T_0$ and $\N_0$ given by \eqref{eq:H0N0withoutbounds}. Then we have the following:
\begin{enumerate}[label=(\roman*)]
\item \label{claimFactcut1} $\forall t \in \T_0, \forall n \notin \N_0, \ y^{\infty}\ih \geq \ubx\ih$  and $ \x^{\infty}\ih = \ubx\ih$ ; 
\item  \label{claimFactcut4} $\forall n \in \N_0, \lambda_n  < 0$ ;
\item \label{claimFactcut3}$\forall t \notin \T_0, \forall n \in \N_0, \ \x^{\infty}\ih =\lbx\ih$ \ ;
\item \label{claimFactcutNonemptysets} the sets $\T_0, \T_0^c$, $\N_0$ and $\N_0^c$ are nonempty.
\end{enumerate}
\end{proposition}

\todo[inline,disable]{SG: As we have almost one page left, I suggest to include
  an appendix, with some hints of the proof (state some lemma). This appendix will be for the referees only, we have no interest in keeping it in the CDC paper - the proofs should be only in the journal. It may be safer however to include such an appendix, otherwise the referees will write that they were unable to check the results, and this may downgrade the paper. \\PJ: Agreed, I'll add that}

The proof of \Cref{prop:factsOncut} relies on the KKT optimality conditions associated to \eqref{pb:QPxE} and \eqref{pb:QPxP}.
 Then we use \Cref{prop:factsOncut} and the convergence condition to prove \Cref{thm:violatedCut}:
\begin{align*}
& \sum_{n\in\N_0} E_n + \sum_{t\in\H_0, n \notin \N_0} \ubx\ih- \sum_{t \notin \T_{0}, n \in \N_{0}} \lbx\ih  -\sum_{t\in\H_0} p_t \\
&=\sum_{t\in\H_0} \left(  \sum_{n\in\N} \x^\infty\ih - \sum_{n\in\N} \y^\infty\ih \right) =\sum_{t\in\H_0} \left(  \sum_{n\in\N} -\nu^\infty_t \right)\\
&=\sum_{t\in\H_0} \left(  - \sum_{n\in\N} \left| \x^\infty\ih - \y^\infty\ih \right|\right)= -\frac{\norm{\xEinf-\xPinf}_1}{2} <0 \ ,
\end{align*}
where the last equality comes from $\sum_{t\in\H_0} \sum_{n\in\N} ( {\x_E}\ih - {\x_P}\ih) = 0$. %
The compact form \eqref{eq:cutConfidWithoutN0} also follow from \Cref{prop:factsOncut}:
\begin{align*}
A_{\T_0}(\xE) & = \sum_{n\in\N_0} E_n +  \sum_{t \in \T_0, n \notin \N_0} \ubx\ih  - \sum_{t \notin \T_0, n \in \N_0} \lbx\ih  \\
&=  \sum_{n\in\N_0, t\in\H_0} \x^\infty\ih + \sum_{t \in \T_0, n \notin \N_0} \x^\infty\ih   = \sum_{t\in\H_0} \sum_{n\in\N} \x^\infty\ih \ .
\end{align*}

\section{Proof of \Cref{thm:cvgAP}}
\label{app:proofCvgRate}
\newcommand{\G}{\mc{G}}
\newcommand{\E}{\mc{E}}
\newcommand{\Lsv}{\Lambda_{\rm{sv}}}
\newcommand{\nul}{\rm{Ker}}
For this analysis, we use the space $\rr^{NT}= \rr^T \times \dots \times \rr^T$, where the $(n-1)T+1$ to $n T$ coordinates correspond to agent $n$, for $1\leq n \leq N$. 
We make use of the following results: 
\begin{lemma}[\cite{nishihara2014convergence}] For APM on polyhedra $\X$ and $\Y$, the sequences $(\xx^{(k})_k$ and $(\yy^{(k})_k$ converge at a  geometric rate, where the rate is bounded by the maximal value of the square of the cosine of the Friedrichs  angle $c_F(U,V)$ between a face $U$ of $\X$ and a face $V$ of $\Y$, where $c_F(U,V)$ is given by:
\begin{equation*}
\begin{split}
c_F(U,V)= \sup \{ & u^T v \ | \norm{u} \leq 1, \norm{v} \leq 1 \\ 
 &  u \in U \cap (U \cap V)^\perp, v \in V \cap (U \cap V)^\perp  \}.
\end{split}
\end{equation*}
\end{lemma}

\begin{lemma}[\cite{nishihara2014convergence}]
Let $A$ and $B$ be matrices with orthonormal rows and with equal numbers of columns and $\Lsv(AB^\tr)$ the set of singular values of $AB^\tr$. Then if $\Lsv(AB^\tr)=\{1\}$, then $c_F (\nul(A), \nul(B )) = 0$. Otherwise,
$c_F (\nul(A), \nul(B))= \max_{\lambda<1}\{\lambda \in \Lsv(AB^\tr) \}$.
\end{lemma}

In our case, the polyhedra $\Y$ is an affine subspace $\Y=\{ \xx \in \rr^{NT} \ | A\xx= \sqrt{N}^{-1} \bm{1}_T \}$ with 
\setcounter{MaxMatrixCols}{20}
$A\eqDef  \sqrt{N}^{-1} J_{1,N} \otimes I_T$,
where $\otimes$ denotes the Kronecker product. The matrix $A$ has orthonormal rows and the direction of $\Y$ is $\text{Ker} (A)$.

Describing the faces $\X$ is more complex. We have a polyhedral description of $\X$, and the faces of $\X$ are subsets of the collection of affine subspaces indexed by $\ (\ol{\T}_n,\ul{\T}_n)_n \subset \T^N$ (with $\ol{\T} \cap \ul{\T} = \emptyset$):
\begin{equation*}
\begin{split}
 \mathcal{A}_{(\ol{\T}_n,\ul{\T}_n)_n} & \eqDef \Big\{ (\xx)_{nt} \ | \forall n, \ \xx_n^\tr \dsoneT =E_n \text{ and } \\
& \forall t \hh\in \ol{\T}_n, \x\ih\hh=\lbx\ih,   \text{ and } \forall t \in \ul{\T}_n, \x\ih\hh=\ubx\ih \Big\}.
\end{split}
\end{equation*}
The associated linear subspace is given by $ \nul(B) $, where the $N$ first rows of $B$ are given by
$[B]_{[N]} \eqDef  \sqrt{T}^{-1} I_N \otimes J_{1,T}$,
and the matrix $B$ has $b\eqDef \sum_n \text{ card} (\T_n ) $ more rows, where $\T_n \eqDef \ul{\T}_n \cup \ol{\T}_n$, corresponding to the saturated inequalities $(\x\ih= \ubx\ih$ or $\lbx\ih$).  
\newcommand{\oneT}{\bm{1}}
\newcommand{\knor}{\kappa}
We denote by $K_n \eqDef \mr{card}(\T_n)$. Then, renormalizing $B$, we can show that the double product $S:= (A B^\tr) (A^\tr B) $, of size $T\times T$ is given by:
\begin{align*}
S\eqDef\frac{1}{N} \begin{pmatrix}
\dssum_n \dfrac{\dsone_{ \{k,\ell\} \subset \T_n^c }}{ T-K_n }
\end{pmatrix}_{ \hh\hh  k,\ell} \hh\hh\hh\hh\hh + \frac{1}{N} \hh \sum_{1 \leq t \leq T} \left(\txt\sum_n \hh \dsone_{t\in\T_n}\right) E_{t,t}.
\end{align*}
Denote $\bar{\T} \eqd \cup_n \T_n^c$ and $P=\eqd I_T-S$. As $P$ can be written as a block diagonal matrix $P=  \text{diag}(P_{\bar{\T}}, 0_{\bar{\T}^c} )$, we can restrict ourselves to the subspace Vect$(e_t)_{t\in\bar{\T}}$ to find the least  positive eigenvalue of $P$, that we denote by $\lambda_1$.

Consider the weighted graph $\G=(\bar{\T},\E)$ whose vertices are the time periods $\bar{\T}$ and edge $(k,\ell)$ has weight $S_{k,\ell}= \tfrac{1}{N}\sum_n \tfrac{\dsone_{ \{k,\ell\} \subset \T_n^c }}{ T-K_n }$ (if this quantity is zero, then there is no edge between $k$ and $\ell$). One can show that $\sum_{\ell \neq k }  -P_{k,\ell} = P_{kk}$,
which shows that $P$ is the Laplacian matrix of $\G$. 

\newcommand{\tm}{\text{-}}
Using the Laplacian property and Cauchy-Schwartz, one shows that for any $u \perp \oneT$:
\begin{align*}
& u^\tr P u   \geq   \min_{k,\ell \in (s^*\text{-}t^*)} (-P_{k,\ell})  \tfrac{(u_{t^*}- u_{s^*})^2 }{ d_{s^*,t^*}}  \geq \tfrac{4T \norm{u}_2^2 }{N(T+1)^2(T-1)^2}  %
\end{align*}
where $u_{t^*}:= \max_t u_t$, $u_{s^*}:= \min_t u_t$  and $d_{s^*,t^*}$ is the the distance between $s^*$ and $t^*$ in $\G$, and $(s^*\tm t^*)$ a path from $s^*$ to $t^*$.

As $\oneT$ is an eigenvector of $P$ associated to $\lambda_0=0$, from the minmax theorem, we get
 $\lambda_1(P)\geq \frac{4}{N(T+1)^2(T-1)} := 1- \rhoN$ and the greatest singular value lower than one of $BA^\tr AB ^\tr$ is $\rhoN$ and then, applying the preceding Lemmas of \cite{nishihara2014convergence}, we obtain the result stated in \Cref{th:boundCvgRate}

\end{appendices}
\end{document}